\documentclass[trackchanges,twocolumn]{aastex7}

\usepackage{subfigure}
\usepackage{comment}
\usepackage{booktabs}

\begin{document}

\title{Revealing the Origins of Galactic Globular Clusters via Their Mg-Al Abundances}

\author[orcid=0009-0006-5216-7648,gname=Shihui, sname='Lin']{Shihui Lin}
\affiliation{School of Physics and Astronomy, Sun Yat-sen University, Zhuhai 519082, China}
\affiliation{CSST Science Center for the Guangdong–Hong Kong–Macau Greater Bay Area, Zhuhai, 519082, China}
\email[]{hui941187@gmail.com}  

\author[orcid=0000-0002-0066-0346,gname=Baitian, sname='Tang']{Baitian Tang} 
\affiliation{School of Physics and Astronomy, Sun Yat-sen University, Zhuhai 519082, China}
\affiliation{CSST Science Center for the Guangdong–Hong Kong–Macau Greater Bay Area, Zhuhai, 519082, China}
\email[show]{tangbt@sysu.edu.cn}

\author[orcid=0009-0007-4049-127X,gname=Genghao, sname='Liu']{Genghao Liu}
\affiliation{School of Physics and Astronomy, Sun Yat-sen University, Zhuhai 519082, China}
\affiliation{CSST Science Center for the Guangdong–Hong Kong–Macau Greater Bay Area, Zhuhai, 519082, China}
\email{liugenghao0310@gmail.com}

\author[orcid=0000-0003-3526-5052]{Jos\'e G. Fern\'andez-Trincado}
\affiliation{Instituto de Astronom\'ia, Universidad Cat\'olica del Norte, Av. Angamos 0610, Antofagasta, Chile}
\email{jose.fernandez@ucn.cl}

\author[orcid=0000-0002-3900-8208]{Douglas Geisler}
\affiliation{Departamento de Astronom\'{i}a, Casilla 160-C, Universidad de Concepci\'{o}n, Concepci\'{o}n, Chile}
\affiliation{Departamento de Astronom\'ia, Facultad de Ciencias, Universidad de La Serena. Av. Ra\'ul Bitr\'an 1305, La Serena, Chile}
\email{none}

\author[orcid=0000-0003-1388-5525]{Guy Worthey}
\affiliation{Department of Physics and Astronomy, Washington State University, Pullman, WA 99163-2814, USA}
\email{none}

\author[orcid=0000-0002-7064-099X]{Dante Minniti}
\affiliation{Depto. de Cs. Físicas, Facultad de Ciencias Exactas, Universidad Andrés Bello, Av. Fernández Concha 700, Las Condes, Santiago, Chile}
\affiliation{Vatican Observatory, V00120 Vatican City State, Italy}
\email{none}






\begin{abstract}

Many Galactic globular clusters (GCs) originated in diverse host galaxies before being subsequently incorporated into the Milky Way through hierarchical galaxy assembly. Identifying their origins is crucial for revealing galaxy properties at early times. Traditional classification methods relying on dynamical properties face inherent uncertainties stemming from the evolving Galactic potential and complex merger histories. Chemically driven classification confronts a distinct obstacle: multiple populations --- abundance variations in light elements of GC members. In this Letter, we identify primordial populations exhibiting lower [Al/Fe] as reliable tracers of their birth environments' chemical evolution. A clear chemical dichotomy emerges between in-situ and accreted GC populations at [Fe/H$]>-1.5$, particularly in the [Mg/Fe]-[Al/Fe] plane, indicating that their progenitor galaxies have experienced fundamentally different enrichment histories. While our chemically driven classification demonstrates general consistency with dynamically driven classifications, notable discrepancies emerge: NGC 288 and M4 are reclassified as in-situ, and Terzan 9 as accreted. This chemically driven GC classification provides promising application for Galactic archaeology.

\end{abstract}



\section{Introduction} 

According to the Lambda cold dark matter model, galaxies in our Universe formed hierarchically. Our Milky Way (MW) aligns with this cosmological framework, having undergone successive mergers and accretions of smaller dwarf galaxies \citep{ibata_dwarf_1994, unavane_merging_1996}.  Globular clusters (GCs), among the oldest stellar systems in galaxies, exhibit ages comparable to their host galaxies \citep{beasley_formation_2002, brodie_extragalactic_2006}. While dwarf galaxies were predominantly disrupted during merger events, their high-density GCs could survive these interactions. Consequently, the present-day Galactic GC system consists of clusters originating from various progenitor galaxies. \citep{forbes_accreted_2010, massari_origin_2019, chen_galaxy_2024}. These ancient stellar systems therefore serve as valuable tracers for reconstructing galactic evolution and merger histories.


If the progenitor galaxy of each GC can be identified, we can decode the detailed evolution of individual progenitor galaxies through their age-metallicity relation \citep[e.g.,][]{Law2010}, or chemical evolution. Thanks to the precise proper motions of GCs provided by the $Gaia$ mission, the clustering in their orbital parameter space links present-day Galactic GCs to their possible progenitor galaxies \citep[e.g.,][]{Callingham2022}. In a recent update of \citet{massari_origin_2019} based on $Gaia$ eDR3 data (hereafter MKH),  Galactic GCs are linked to 10 progenitors: main bulge, main disk, Sagittarius (Sag), Helmi stream (H99), Gaia-Enceladus (GE), Sequoia (Seq), Low-energy group, Elqui, Cetus, and High-energy group. Only the first two groups are considered in-situ, while other groups are classified as accreted.

However, the assumption of invariant orbital parameters becomes tenuous when considering the Galaxy's complex merger history and long-term dynamical evolution. Simulations that include realistic ISM prescriptions also suggest that dynamical criteria alone may be problematic \citep[e.g.,][]{Pagnini2023}. For instance, based on chemical similarities to other in-situ GCs, \citet{ceccarelli_cluster_2025} argued that NGC 288 formed within the Galactic proto-disk, and was later dynamically heated during the GE merger. 
Unlike dynamical properties, chemical abundances remain largely unaffected by galaxy evolution or merger events. Consequently, they serve as robust tracers for identifying the origins of Galactic field stars. The [Mg/Mn]-[Al/Fe] plane is widely used for this purpose \citep[e.g.,][]{hawkins_using_2015, das_ages_2020}, particularly as \citet{horta_evidence_2021} demonstrated distinct evolutionary tracks for MW-like and GE-like galaxies within it, providing a theoretical basis for empirical division lines. However, categorizing early formed metal-poor stars remains challenging, as they predominantly occupy overlapping regions (specifically the upper left) in this plane.

Chemically tagging the origins of Galactic GCs faces a further complication: multiple populations (MPs). MPs alter light-element abundances (e.g., C, N, O, Na, Mg, Al, Si) within metal-poor/compact stellar systems \citep{Carretta2009, Tang2017, meszaros_homogeneous_2020, HuangTang2024}. A GC forms from a natal cloud inheriting its parent galaxy's chemistry. However, after massive stars ($\sim 8-1\times10^4 M_{\odot}$) are born, their strong winds eject processed material. This enriched gas mixes with primordial gas, altering the composition of subsequently formed stars. Retaining this enriched gas against expulsion by stellar feedback requires a compact gas/stellar environment \citep{Krause2016, HuangTang2024}. To mitigate MP effects when chemically tagging the Galactic GCs, studies typically employ average chemical abundances. For example, \citet{horta_chemical_2020} distinguished in-situ from accreted origins among relatively metal-rich GCs ([Fe/H$]>-1.5$) using mean [Si/Fe]. Conversely, \citet{belokurov_-situ_2024} showed that the mean [Mg/Fe]-[Al/Fe] plane fails to clearly separate in-situ and accreted GCs defined by their $E-L_z$ (orbital energy vs. vertical angular momentum) location. This difficulty arises because Mg, Al, and Si exhibit significant variations within individual GCs \citep[e.g.,][]{Tang2018, Milone2022}, complicating classification in certain cases.

In this Letter, we aim to disentangle the distinct chemical abundance contributions from a GC's progenitor galaxy and its internal MP enrichment. Crucially, primordial populations genuinely reflect the progenitor galaxy's chemistry, enabling a $bona~fide$, chemically driven classification of GCs. 
This Letter is structured as follows. Section 2 details our selection criteria for GC member stars. Section 3 presents our GC classification results based on their locations in the [Al/Fe]-[Fe/H] and [Mg/Fe]-[Al/Fe] planes. In Section 4, we discuss the physics behind and robustness of this classification; we further expand our GC sample to the latest data release. Section 5 provides a brief summary and outlines future perspectives.


\section{DATA} 

The Apache Point Observatory Galactic Evolution Experiment (APOGEE, \citealt{{Majewski2017}}) delivers high-resolution ($R\sim$22,500) $H$-band spectra ($\lambda = 1.51 - 1.69$ $\mu$m) from both the SDSS 2.5 m telescope at Apache Point Observatory \citep{Gunn2006} and the 2.5 m du Pont telescope at Las Campanas Observatory \citep{Bowen1973}. APOGEE data reduction software was applied to reduce multiple 3D raw data cubes into calibrated, well-sampled, combined 1D spectra \citep{Nidever2015}. When observing GC stars, targets were selected and prioritized using a combination of preexisting information (i.e., stellar parameters, abundances, radial velocities, proper motions, and location in the color-magnitude diagram) and observing constraints imposed by APOGEE magnitude and fiber collision limits \citep{Zasowski2017}. Such target selection is insensitive to MP.
Given the typically weaker spectral lines at low metallicity (e.g., [Fe/H$]=-2$) and possible large variations of elemental feature lines (e.g., Al) in GC members, a line-by-line analysis for individual chemical species is preferred (e.g., Brussels Automatic Code for Characterizing High-accuracy Spectra, BACCHUS, \citealt{Masseron2016}) over abundance analysis based on a set of spectral windows (e.g., APOGEE Stellar Parameter and Chemical Abundance Pipeline, ASPCAP, \citealt{jonsson_apogee_2020}).
We therefore adopt the homogeneous chemical abundances derived by \citet{meszaros_homogeneous_2020} using BACCHUS for 2283 red giant stars in 31 Galactic GCs. These authors select GC members based on their similarities in radial velocities, metallicities, and spatial locations (i.e., inside the GC tidal radius). We selected accurate Fe, Mg, Al, and Si abundances meeting the following criteria: (1) spectral signal-to-noise ratio (SNR) $>70$; (2) effective temperature ($T_{\mathrm{eff}}$) $<5500$ K; and (3) abundance uncertainties $<$ 0.2 dex. Additionally, we require clusters to have at least six member stars to ensure statistically robust separation of primordial and enriched populations in subsequent analysis. Our final sample comprises 27 Galactic GCs with 2,142 stars (Table \ref{tab:M20}).

\begin{figure*}[ht!]
\centering
\includegraphics[width=0.7\textwidth]{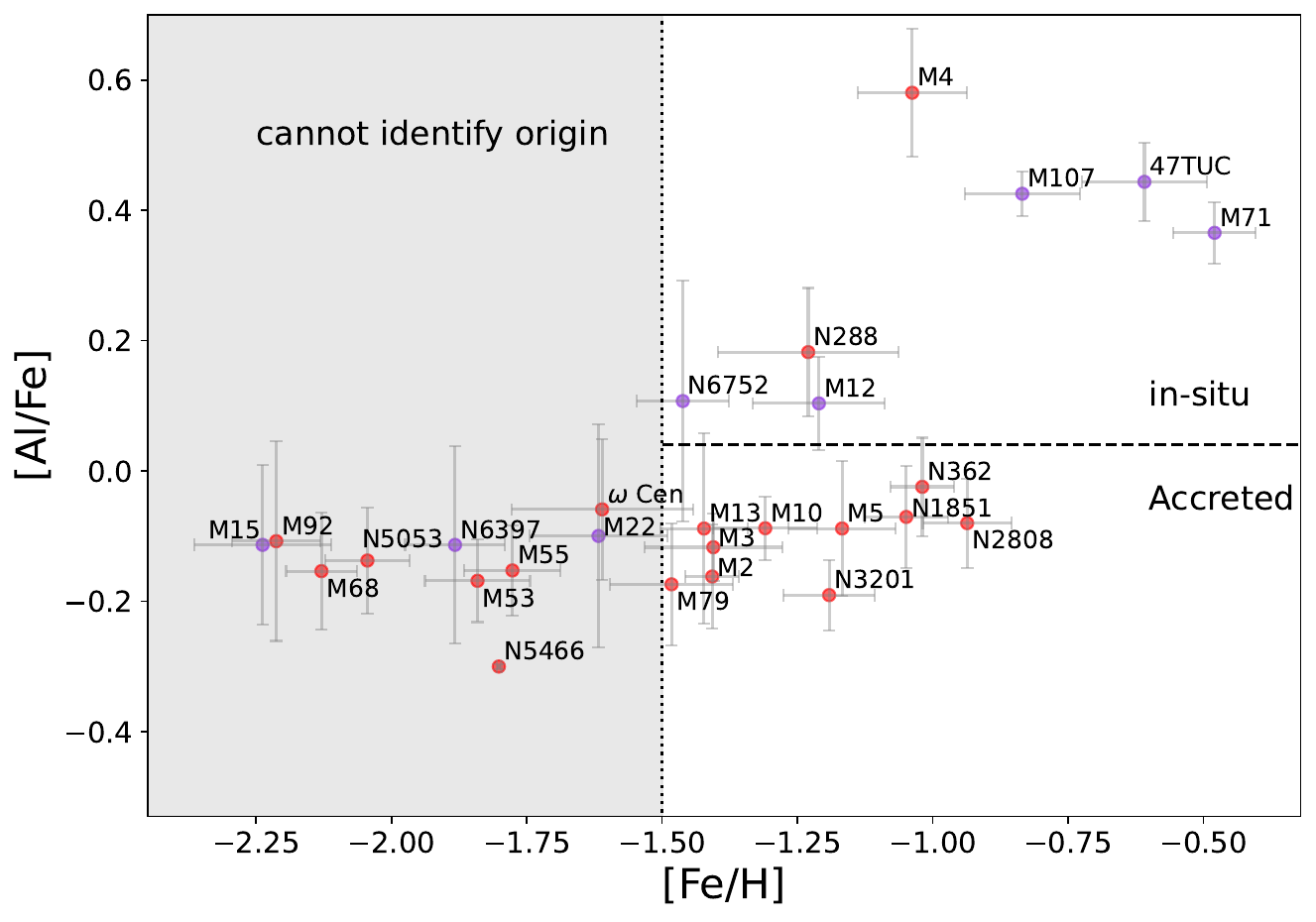}
\caption{Mean [Al/Fe] versus mean [Fe/H] of primordial populations in Galactic GCs. Their associated standard deviations are shown as error bars. 
The black dotted line indicates the metallicity of [Fe/H$]=-1.5$. GCs with [Fe/H$]<-1.5$ are not suitable for our chemically based classification (grey region). The black dashed line separates chemically classified in-situ and accreted clusters. In comparison, MKH classification is also shown (in-situ: purple, accreted: red).
\label{fig:Al-Fe}}
\end{figure*}

\begin{figure}[ht!]
\centering
\includegraphics[width=0.45\textwidth]{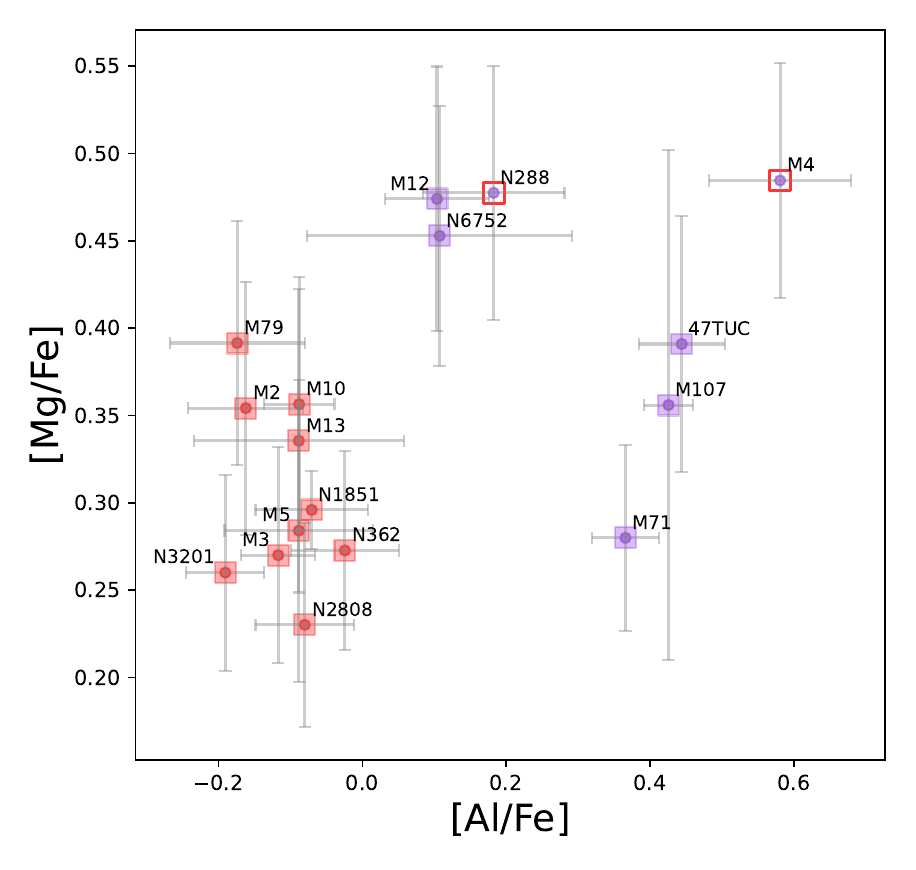}
\caption{Mean [Mg/Fe] versus mean [Al/Fe] of primordial populations in Galactic GCs. Their associated standard deviations
are shown as error bars. Small purple (red) dots represent the in-situ (accreted) GCs based on our classification. Filled purple (red) squares mark the in-situ (accreted) GCs based on MKH classification. Two GCs (NGC 288 and M4) with inconsistent classification between the two works are properly labeled. 
}
\label{fig:Mg-Al}
\end{figure}

\section{Results} 
\label{sec:results}

\begin{figure*}[ht!]
\centering
\includegraphics[width=0.85\textwidth]{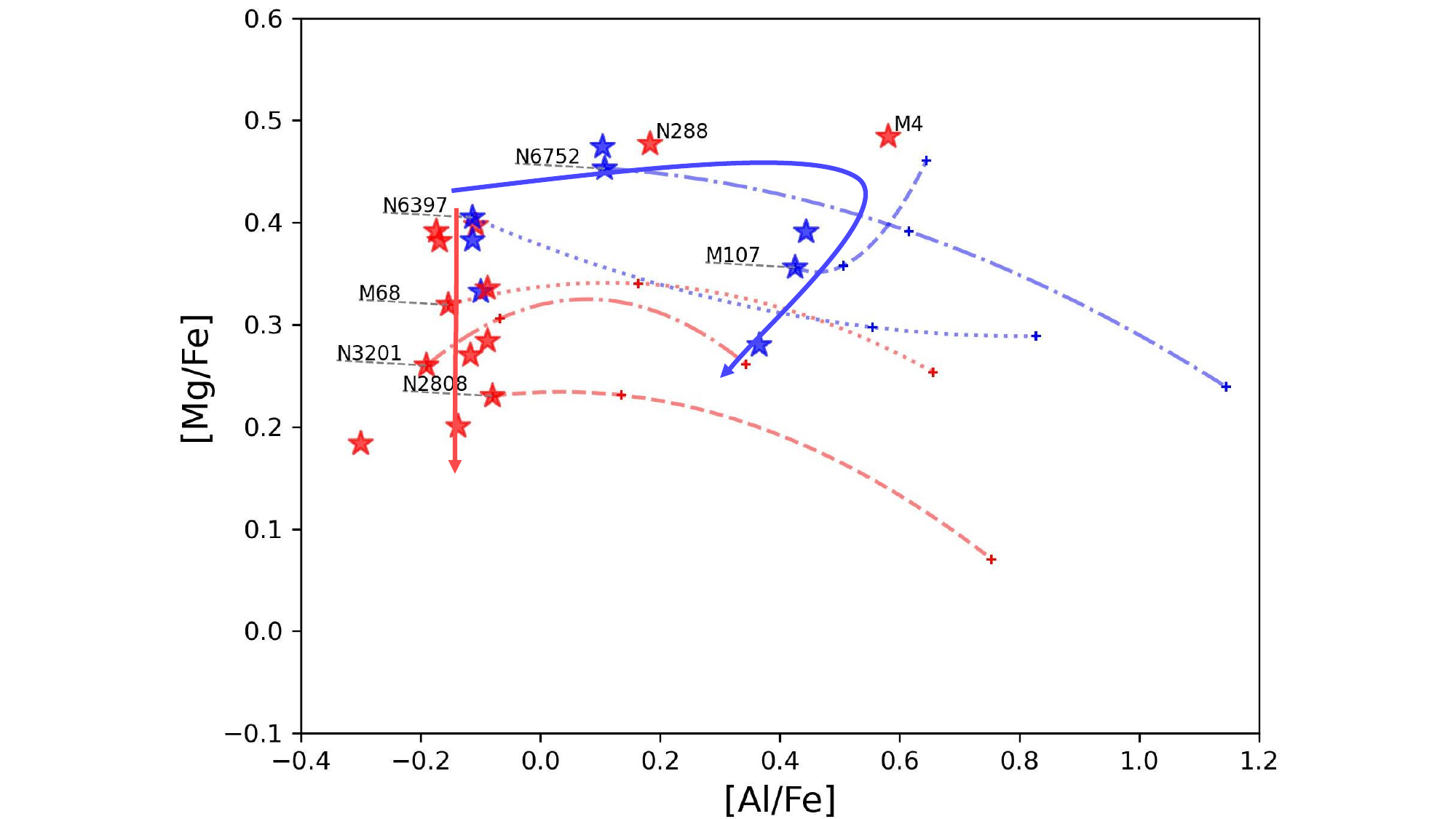}
\caption{Our conceptual diagram showing mean [Mg/Fe] versus mean [Al/Fe] of primordial populations in Galactic GCs. The Blue (red) star symbols mark the in-situ (accreted) GCs classified by MKH. The blue (red) arrowed line shows the empirical evolutionary path of in-situ (accreted) GC primordial populations, with increasing metallicity. The Mg-Al (anti-)correlations of in-situ GCs are shown as: blue dotted line - NGC 6397 ([Fe/H$]=-1.88$), blue dash-dotted line - NGC 6752 ([Fe/H$]=-1.46$), blue dashed line - M107 ([Fe/H$]=-0.85$), respectively. Meanwhile, The Mg-Al (anti-)correlations of accreted GCs are shown as: red dotted line - M86 ([Fe/H$]=-2.16$), red dash-dotted line - NGC 3201 ([Fe/H$]=-1.22$), red dashed line - NGC 2808 ([Fe/H$]=-0.93$), respectively.
\label{fig:general}}
\end{figure*}

Among the elements analyzed, Al abundances exhibit the most significant variations, making them ideal for population separation. Since Al is strongly enhanced by GC enrichment, revealing the progenitor galaxy's chemistry necessitates identifying the less Al-enhanced subpopulation. In line with the literature, we term this subpopulation ``primordial populations'' (or ``first generation stars''), adopting slightly different definition (see Section \ref{sec:prim} for more discussions).  Specifically, we define them as follows. Stars within each GC are sorted by their [Al/Fe] in ascending order. The primordial population is defined as the first $N/n$ stars, where $N$ is the sample size of the GC members (Table \ref{tab:M20}) and $n$ is the fixed number of subpopulations assigned to each GC. We apply the same $n$ value for all studied GCs. When $N/n$ is non-integer, we use its floor value (greatest integer less than or equal to $N/n$). As $n$ increases, both the subsample size and mean [Al/Fe] value of the primordial population decrease. We adopt $n=3$ for subsequent analysis and further discuss this choice in Section \ref{sec:prim}. 

At [Fe/H$]>-1.5$, a clear separation emerges (at [Al/Fe$] \sim 0 $) between in-situ ([Al/Fe$] > 0.04$ dex) and accreted GCs ([Al/Fe$] < 0.04$ dex) in the [Al/Fe]-[Fe/H] plane (Figure \ref{fig:Al-Fe}). 
These two groups also occupy distinct regions in the [Mg/Fe]-[Al/Fe] plane (Figure \ref{fig:Mg-Al}). We therefore focus our origin analysis on relatively metal-rich GCs. The metallicity limit ([Fe/H$]=-1.5$) will be further discussed in Section \ref{sect:phy}.
To validate our classification, we compare it with the dynamically driven MKH classification. The agreement is generally good, except for NGC 288 and M4. Although classified as accreted by MKH, their primordial populations exhibit elevated [Al/Fe] abundances comparable to other in-situ GCs. As noted in Section 1, \citet{ceccarelli_cluster_2025} proposed an in-situ origin for NGC 288, suggesting subsequent dynamical heating. Similarly, M4 was classified as in-situ based on its mean [Si/Fe] \citep{horta_chemical_2020}. Given that chemical abundances remain unaffected by changes in the galactic potential, we conclude that our chemically based classification provides a more robust indicator of GC origin. See Section \ref{sect:phy} for more discussions about their implications.

The [Mg/Fe]-[Al/Fe] plane is frequently used to investigate MPs in GCs \citep{pancino_gaia_2017,meszaros_homogeneous_2020}. In this Letter, we propose a novel framework to disentangle galactic chemical enrichment from GC self-enrichment, illustrated conceptually in Figure \ref{fig:general}. This separation is achieved by analyzing chemical abundances exclusively within GC primordial populations. We first exclude GCs exhibiting significant iron spreads, (e.g., $\omega$ Cen, which shows complex star formation histories; \citealt{Jofre2025, Mason2025}). Our analysis reveals distinct evolutionary pathways for in-situ and accreted GCs: 
\begin{itemize}
\item {\bf Low metallicity GCs [Fe/H$]<-1.5$:} Both in-situ and accreted GCs occupy a similar region ([Mg/Fe$]\sim 0.3-0.4$ and [Al/Fe$]\sim 0.1$), indicating indistinguishable chemical evolution between the early MW and its accreted dwarf galaxies at these metallicities. 
\item  {\bf Accreted GCs ([Fe/H$]>-1.5$):} Progenitor galaxies show a pronounced decrease in [Mg/Fe] but only a mild decrease in [Al/Fe] with increasing metallicity. 
\item  {\bf In-situ GCs ([Fe/H$]>-1.5$):} The MW progenitor exhibits more complex evolution. Between [Fe/H$]=-1.5$ and $-1.0$, [Mg/Fe] remains nearly constant while [Al/Fe] rises significantly to $\sim 0.5$. Notably, NGC 288 and M4 within this metallicity range are chemically classified as in-situ. Above [Fe/H$]=-1.0$, both [Mg/Fe] and [Al/Fe] decrease with metallicity. 
\end{itemize}
Crucially, the chemical evolution patterns revealed by GC primordial populations align remarkably with those observed in field stars of their progenitor galaxies \citep[e.g.,][]{Tang2023}. The physics behind measured chemical abundances is further discussed in Section \ref{sect:phy}.

To illustrate the complexity of the [Mg/Fe]-[Al/Fe] plane for GC member stars, Figure \ref{fig:general} presents Mg-Al (anti-)correlations for three representative in-situ GCs (M68, NGC 3201, NGC 2808), and three accreted GCs (NGC 6397, NGC 6752, M107). These GCs range from metal poor to metal rich (Figure \ref{fig:Al-Fe}). Al variations in GCs depend on both metallicity and cluster mass \citep{pancino_gaia_2017, meszaros_homogeneous_2020}.


\section{Discussion}

\subsection{Primordial Population}
\label{sec:prim}

Our definition of the primordial population differs from that based on HST photometry, which primarily relies on the UV NH molecular band \citep{Milone2022}. We instead define it using derived Al abundances, which exhibit a continuous distribution without clear boundary between primordial and enriched populations. Consequently, our primordial population refers to ``stars not strongly enriched in Al by the polluters responsible for GC MP''. 

The primordial population is defined as the first $N/n$ stars with lower [Al/Fe], and we adopt $n=3$ in Section \ref{sec:results}. 
In principle, a larger $n$ yields a less Al-enhanced subpopulation, thereby better reflecting the true progenitor galaxy's chemistry. However, smaller subsamples are prone to statistical fluctuations and are less viable for GCs with few confirmed members.
Testing values of $n=3,4,6,8$ revealed that the [Al/Fe] distinction between in-situ and accreted GCs at [Fe/H$]>-1.5$ is most pronounced when $n=3$ (Figure \ref{fig:Al-Fe}). This choice also provides comparatively larger sample sizes of primordial populations, reducing statistical fluctuations in their mean abundances. 
We therefore adopt $n=3$ in this Letter.

Since APOGEE targeting is insensitive to MP, one might be concerned that all observed stars in a given GC could belong to the enriched population, leaving no  primordial stars. The probability ($p$) of this occurring is $(\frac{1}{2})^N$, where N is the number of stars observed in the GC. For Our minimum sample size (N=6), $p=0.0156$. This probability quickly drops to $<10^{-3}$ for $N\geqslant 10$.


[Al/Fe$]=0.3$ has been proposed to separate primordial and enriched populations in GCs when Na lines are inaccessible \citep[e.g.,][]{meszaros_homogeneous_2020}. However, the clear distinction in primordial [Al/Fe] between in-situ and accreted GCs at [Fe/H$]>-1.5$ necessitates caution. Applying this fixed threshold risks underestimating primordial populations, particularly in in-situ systems where primordial [Al/Fe] is typically higher.

\subsection{Underlying Physics}
\label{sect:phy}

Theoretical models by \citet{horta_evidence_2021} indicate that the chemical evolution tracks of MW-like and GE-like galaxies in the [Mg/Mn]-[Al/Fe] plane are clearly separated at [Fe/H$]=-0.8$ (black crosses in their Figure 2). However, these two tracks start diverging at lower metallicity, specifically during the $0.3–1$ Gyr evolution phase of the GE-like system. Unfortunately, precise metallicities along these tracks cannot be robustly constrained. Based on GC's mean [Si/Fe] distribution \citep{horta_chemical_2020} and primordial [Al/Fe] distribution (our Figure \ref{fig:Al-Fe}), in-situ and accreted GCs are observationally indistinguishable below [Fe/H$]=-1.5$. Interestingly, such limit may also be identified in other $\alpha$-elements. For example, \citet{horta2025} derive the hex ratios for Galactic GCs --- ratios between hydrostatic $\alpha$-elements (namely, ([Mg/Fe]+[O/Fe])/2) and explosive $\alpha$-elements (namely, ([Si/Fe]+[Ca/Fe]+[Ti/Fe])/3).  In-situ and accreted GCs are also indistinguishable below [Fe/H$]=-1.5$, but in-situ GCs generally show larger hex ratios than accreted GCs at [Fe/H$] \gtrsim -1.5$ (their Figure 4).

Current theoretical models explain the different evolution tracks between MW (thick disk and in-situ halo) and its accreted dwarf galaxies at [Fe/H$]>-1.5$ as follows: The MW's deep potential well draws large amounts of gas, sustaining a high star formation rate and rapid enrichment. This increases the fraction of massive stars formed \citep{Jerabkova2018}, which produce large yields of C, N, O, Ne, and Mg due to their massive mantles \citep{K06}.  The higher fraction of massive stars is further confirmed by their hex ratios: in-situ GCs generally show larger hex ratios than accreted GCs at [Fe/H$] \gtrsim -1.5$ (Figure 4 of \citealt{horta2025}). Higher hex ratios indicate an initial mass function (IMF) with a larger amount of most massive stars \citep{Carlin2018,McWilliam2013}.
As one of the odd-Z elements, Al is enhanced by the surplus of neutrons in $^{22}$Ne, and $^{22}$Ne is transformed from $^{14}$N by the CNO cycle during He burning. Consequently, Al yield correlates with the abundances of CNO elements, making it metallicity-dependent \citep[e.g., Fig. 5 of][]{K06}.  Conversely, due to slow chemical enrichment in GE-like galaxies, Type Ia supernova (SNe Ia) --- which produce substantial amounts of Fe --- begin to significantly influence galactic chemical evolution at comparatively low metallicities ([Fe/H$] \sim -1.5$).
Given the distinct evolutionary pathways of Mg, Al, and Fe, we expect higher [Mg/Fe] and [Al/Fe] ratios in MW in-situ stars than in accreted halo stars ([Fe/H$]>-1.5$), matching observations by \citet{hawkins_using_2015}.

Recently, \citet{xiang2022} suggested the MW's old disk began forming $\sim 13$ Gyr ago, only 0.8 Gyr after the Big Bang. Most stars in this component formed about 11 Gyr ago. Consequently, in-situ stars and GCs born in this old (thick) disk may have experienced the ``Splash'' phase \citep{Belokurov2020} induced by the GE merger. In this context, the two misclassified GCs --- NGC 288 and M4 --- may not be coincidental. More ``Splash'' GCs are likely to exist.

\subsection{Robustness}


The [Mg/Fe]-[Al/Fe] plane was also employed to discuss GC origins in \citet{belokurov_-situ_2024}, albeit with a distinct classification methodology. They first established a boundary in $E-L_z$ space separating in-situ and accreted objects using stars with robust [Al/Fe]-based classification ($-1.4<[$Fe/H$]<-1.0$), then classified GCs dynamically based on their $E-L_z$ positions. Their work showed that the mean [Mg/Fe]-[Al/Fe] plane cannot clearly distinguish in-situ from accreted GCs, effectively constituting a $chemically~informed$ dynamical classification. Comparing classifications for GCs with [Fe/H$]>-1.5$, we identify four discrepancies: NGC 288 (classified accreted by Belokurov et al.) versus M13, M10, and NGC 6544 (classified in-situ). As Figure \ref{fig:Al-Fe} shows, the primordial populations of M13, M10, and NGC 6544 exhibit low [Al/Fe] --- a signature typical of less massive dwarf galaxies at [Fe/H$]>-1.5$. Our classification aligns with MKH for these three GCs. 

Given that Si may also vary substantially within GCs, we repeat our classification replacing Mg with Si. The classification results remain unchanged. More importantly, the primordial [Al/Fe] exhibits a significantly larger difference between in-situ and accreted GCs compared to primordial [Si/Fe]. Specifically, $\Delta$[Al/Fe] is $\sim0.25$ dex, while $\Delta$[Si/Fe] is only $\sim0.08$ dex in the range $-1.5<[$Fe/H$]<-1.0$, The [Si/Fe] difference here is consistent with \citet{horta_chemical_2020} and \citet{geisler_capos_2021}. In the higher metallicity range ($-1.0<[$Fe/H$]<-0.7$), $\Delta$[Al/Fe] reaches $\sim0.5$ dex compared to $\Delta$[Si/Fe] $\sim0.15$ dex. These differentials align with the $\Delta$[Al/Fe] and $\Delta$[Si/Fe] values distinguishing in-situ halo and thick disk stars from accreted halo stars reported by \citet{hawkins_using_2015}. Consequently, the enhanced discrepancy in primordial [Al/Fe] provides a more robust basis for classification.

\begin{figure}[ht!]
\centering
\includegraphics[width=0.45\textwidth]{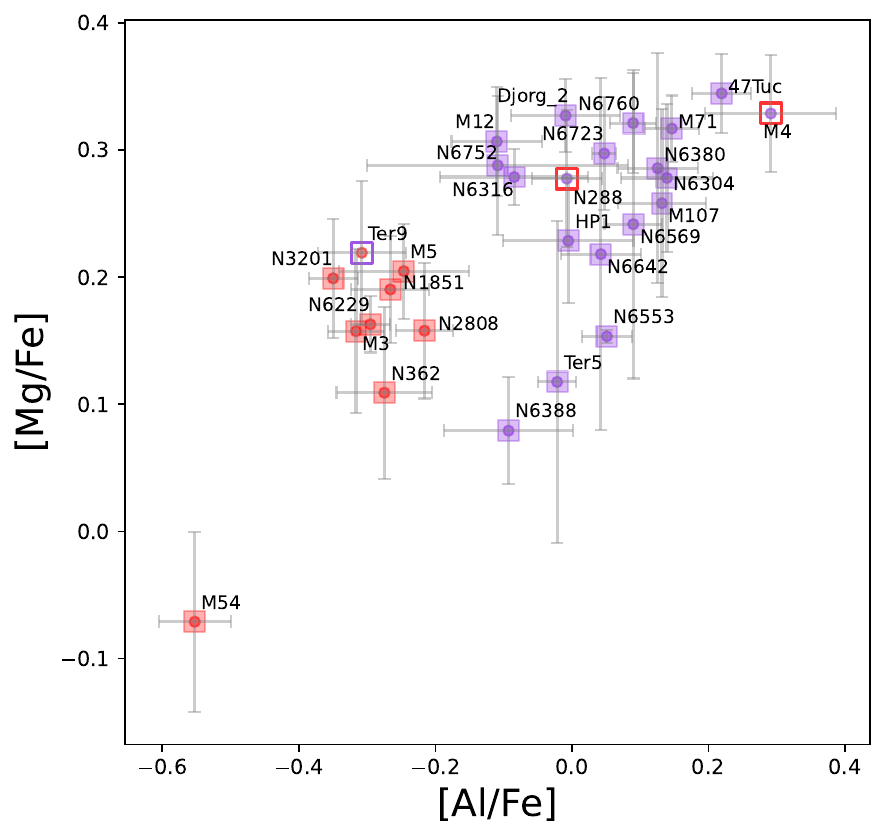}
\caption{Mean [Mg/Fe] versus mean [Al/Fe] of primordial populations in APOGEE GCs. Their associated standard deviations
are shown as error bars. Small purple (red) dots represent the in-situ (accreted) GCs based on our classification. Filled purple (red) squares mark the in-situ (accreted) GCs based on MKH classification. Three GCs (Terzan 9, NGC 288 and M4) with inconsistent classification between the two works are properly labeled. 
}
\label{fig:Mg-Alall}
\end{figure}

\subsection{Expanding the Sample}

Since the homogeneous study by \citet{meszaros_homogeneous_2020}, the APOGEE survey has observed additional GCs, particularly in the bulge region \citep{geisler_capos_2021,FT2021}. The value-added catalog by \citet{schiavon_apogee_2024} provides stellar abundances for 72 Galactic GCs derived through the APOGEE Stellar Parameter and Chemical Abundance Pipeline (ASPCAP, \citealt{jonsson_apogee_2020}). While ASPCAP may struggle with weak spectral features, it remains robust for studying Mg, Al, Si abundances in GCs with [Fe/H$]>-1.5$. We are left with 28 GCs after applying the following criteria: (1) GC mean [Fe/H$]>-1.5$; (2) no STAR\_BAD flag; (3) element flags of Mg, Al, and Si equal zero; (4) sample size larger than 5 (Table \ref{tab:asp}). Figure \ref{fig:Mg-Alall} shows the primordial population abundances for these GCs alongside two classification schemes.
Our chemically driven classification generally agrees with the MKH classification, except for NGC 288, M4 and Terzan 9. The first two GCs have been discussed in Section \ref{sec:results}. Terzan 9, a bulge GC traditionally classified as in-situ \citep[e.g.,][MKH]{ortolani_blue_1999}, exhibits an accreted signature in the [Mg/Fe]–[Al/Fe] plane. This interpretation is reinforced by its Si abundances. Its 11 member stars yield mean [Fe/H$]=-1.37$ and mean [Si/Fe$]=0.18$. For bulge GCs with $-1.5<[$Fe/H$]<-1.0$, both \citet{horta_chemical_2020} and \citet{geisler_capos_2021} report higher mean [Si/Fe] in in-situ GCs ($\sim 0.25-0.26$) versus accreted GCs ($\sim 0.17-0.18$). Terzan 9's low mean [Si/Fe] thus supports an accreted origin. Chemodynamical evidence indicates it formed ex-situ and was later accreted into the bulge. The complex dynamical evolution inside the bulge may change its orbits. In contrast, $bona~fide$ Bulge GCs near mean [Mg/Fe$]\sim 0.13$, [Al/Fe$]\sim 0$ (i.e., NGC 6553, Terzan 5 and NGC 6388) follow the in-situ chemical trend (blue arrow, Figure \ref{fig:general}).

\section{Conclusion}

In this Letter, we present a novel approach to identify GC origins using their Mg-Al distributions. This work is part of our ongoing research initiative, ``Scrutinizing {\bf GA}laxy-{\bf ST}a{\bf R} cluster coevoluti{\bf ON} with chem{\bf O}dyna{\bf MI}cs ({\bf GASTRONOMI})'', which leverages multi-wavelength photometric and spectroscopic data to unravel the coevolutionary relationships between the MW, its satellite dwarf galaxies, and their stellar clusters.
We demonstrate that the chemistry of a GC's progenitor galaxy is encoded within its primordial population. The in-situ and accreted GCs at [Fe/H$]>-1.5$ are clearly distinguished by their primordial [Mg/Fe] and [Al/Fe]. This chemically driven classification is robust against dynamical changes induced by Galactic potential evolution and merger history. In this scheme, NGC 288 and M4 are reclassified as in-situ, while Terzan 9 as accreted. The future of this classification scheme is highly promising, as more metal-rich GCs are identified and observed spectroscopically within the bulge \citep[e.g.,][]{ Minniti2017, Garro2022}.

\begin{acknowledgements}
We thank Jianhui Lian, Jianrong Shi, and Ruoyun Huang for helpful discussions. 
We thank the anonymous referee for insightful comments.
S.L. and B.T. gratefully acknowledge support from the National Natural Science Foundation of China through grants NOs. 12473035 and 12233013, China Manned Space Project under grant NO. CMS-CSST-2025-A13 and CMS-CSST-2021-A08. J.G.F.-T. gratefully acknowledges the grant support provided by ANID Fondecyt Iniciaci\'on No. 11220340, ANID Fondecyt Postdoc No. 3230001, and from the Joint Committee ESO-Government of Chile under the agreement 2023 ORP 062/2023. D.G. gratefully acknowledges the support provided by Fondecyt regular no. 1220264.
D.G. also acknowledges financial support from the Direcci\'on de Investigaci\'on y Desarrollo de la Universidad de La Serena through the Programa de Incentivo a la Investigaci\'on de Acad\'emicos (PIA-DIDULS).
\end{acknowledgements}

\appendix
\begin{deluxetable}{lcccccccc}
\tablewidth{0pt}
\tablecaption{Globular Clusters selected from \citet{meszaros_homogeneous_2020}\label{tab:M20}}
\tablehead{
\colhead{GC name} & \colhead{sample size$^a$} & \colhead{[Fe/H]$^b_{\rm prim}$} & \colhead{$\sigma^c_{\rm Fe}$} & \colhead{[Al/Fe]$^b_{\rm prim}$} & \colhead{$\sigma^c_{\rm Al}$} & \colhead{[Mg/Fe]$^b_{\rm prim}$} & \colhead{$\sigma^c_{\rm Mg}$} & \colhead{MKH classification} 
}
\startdata
\multicolumn{9}{c}{GCs with origins that cannot be identified chemically} \\
\cmidrule{1-9}
M15 & 51(17) & -2.24 & 0.13 & -0.11 & 0.12 & 0.38 & 0.09 & main disk \\
M92 & 34(11) & -2.21 & 0.08 & -0.11 & 0.15 & 0.40 & 0.09 & GE \\
M68 & 25(8) & -2.13 &  0.07 & -0.15 & 0.09 & 0.32 & 0.09 & H99\\
NGC 5053 & 11(3) & -2.04 & 0.08 & -0.14 & 0.08 & 0.20 & 0.09 & H99\\
NGC 6397 & 94(31) & -1.88 & 0.09 & -0.11 & 0.15 & 0.40 & 0.10 & main disk\\
M53 & 31(10) & -1.84 & 0.10 & -0.17 & 0.06 & 0.38 & 0.04& H99\\
NGC 5466 & 7(2) & -1.80 & 0.00 & -0.30 & 0.00 & 0.18 & 0.02& Seq\\
M55 & 54(18) & -1.78 & 0.09 & -0.15 & 0.07 & 0.49 &0.07& Low-energy \\
M22 & 15(5) & -1.62 & 0.13 & -0.10 & 0.17 & 0.33 &0.15 & main disk \\
$\omega$ Cen & 665(221) & -1.61 & 0.17 & -0.06 & 0.11 & 0.46 & 0.11& GE/Seq\\
\cmidrule{1-9}
\multicolumn{9}{c}{accreted GCs} \\
\cmidrule{1-9}
M79 &21(7) & -1.48 & 0.11 & -0.17 & 0.09 & 0.39 & 0.07& GE\\
M13 & 60(20) & -1.42 & 0.08 & -0.09 & 0.15 & 0.34 & 0.09&  GE\\
M2 & 21(7) & -1.41 & 0.05 & -0.16 & 0.08 & 0.35 & 0.07&  GE\\
M3 & 132(44) & -1.41 & 0.13 & -0.12 & 0.05 & 0.27 & 0.06& H99 \\
M10 & 73(24) & -1.31 & 0.10 & -0.09 & 0.05 & 0.36 & 0.07& Low-energy \\
NGC 3201 & 35(11) & -1.19 & 0.08 & -0.19 & 0.05 & 0.26 & 0.06 & Seq/GE \\
M5 & 180(60) & -1.17 & 0.10 & -0.09 & 0.10 & 0.28 & 0.09& GE/H99 \\
NGC 1851 & 27(9) & -1.05 & 0.08 & -0.07 & 0.08 & 0.30 & 0.02&  GE\\
NGC 362 & 37(12) & -1.02 & 0.06 & -0.02 & 0.08 & 0.27 & 0.06& GE \\
NGC 2808 & 57(19) & -0.94 & 0.08 & -0.08 & 0.07 & 0.23 & 0.06& GE \\
\cmidrule{1-9}
\multicolumn{9}{c}{in-situ GCs} \\
\cmidrule{1-9}
NGC 6752 & 124(41) & -1.46 & 0.08 & 0.11 & 0.18 & 0.45 & 0.07&  main disk\\
NGC 288 & 36(12) & -1.23 & 0.17 & 0.18 & 0.10 & 0.48 & 0.07& GE \\
M12 & 49(16) & -1.21 & 0.12 & 0.10 & 0.07 & 0.47 & 0.08&  main disk\\
M4 & 123(41) & -1.04 & 0.10 & 0.58 & 0.10 & 0.48 & 0.07& Low-energy \\
M107 & 41(13) & -0.83 & 0.11 & 0.43 & 0.03 & 0.36 & 0.15& main bulge \\
47 Tuc & 111(37) & -0.61 & 0.12 & 0.44 & 0.06 & 0.39 & 0.07& main disk \\
M71 & 28(9) & -0.48 & 0.08 & 0.37 & 0.05 & 0.28 & 0.05& main disk \\
\enddata
\tablecomments{$^a$: total sample sizes of every GCs, and sample sizes of primordial populations in parentheses. $^b$: mean abundances of primordial populations. $^c$: the associated standard deviations.
}
\end{deluxetable}

\begin{deluxetable}{lcccccccc}
\tablewidth{0pt}
\tablecaption{GCs selected from ASPCAP\label{tab:asp}}
\tablehead{
\colhead{GC name} & \colhead{sample size$^a$} & \colhead{[Fe/H]$^b_{\rm prim}$} & \colhead{$\sigma^c_{\rm Fe}$} & \colhead{[Al/Fe]$^b_{\rm prim}$} & \colhead{$\sigma^c_{\rm Al}$} & \colhead{[Mg/Fe]$^b_{\rm prim}$} & \colhead{$\sigma^c_{\rm Mg}$} & \colhead{MKH classification} 
}
\startdata
\multicolumn{9}{c}{accreted GCs} \\
\cmidrule{1-9}
M3 & 261(87) & -1.48 & 0.11 & -0.32 & 0.04 & 0.16 & 0.06& H99 \\
NGC 3201 & 165(55) & -1.40 & 0.09 & -0.35 & 0.04 & 0.20 & 0.05 & Seq/GE \\
Terzan 9 & 11(3) & -1.37 & 0.10 & -0.31 & 0.06 & 0.22 & 0.06& main bulge \\
NGC 6229 & 8(2) & -1.37 & 0.01 & -0.30 & 0.03 & 0.16 & 0.02& GE \\
M5 & 216(72) & -1.25 & 0.09 & -0.25 & 0.10 & 0.20 & 0.04& GE/H99 \\
NGC 1851 & 49(16) & -1.16 & 0.05 & -0.27 & 0.06 & 0.19 & 0.04&  GE\\
NGC 362 & 65(21) & -1.14 & 0.07 & -0.27 & 0.07 & 0.11 & 0.07& GE \\
NGC 2808 & 124(41) & -1.11 & 0.05 & -0.22 & 0.04 & 0.16 & 0.05& GE \\
M54 & 863(287) & -0.69 & 0.23 & -0.55 & 0.05 & -0.07 & 0.07& Sag \\
\cmidrule{1-9}
\multicolumn{9}{c}{in-situ GCs} \\
\cmidrule{1-9}
NGC 6752 & 138(46) & -1.48 & 0.23 & -0.11 & 0.19 & 0.29 & 0.05&  main disk\\
NGC 288 & 40(13) & -1.33 & 0.06 & -0.01 & 0.05 & 0.28 & 0.05& GE \\
M12 & 78(26) & -1.32 & 0.06 & -0.11 & 0.07 & 0.31 & 0.04&  main disk\\
HP 1 & 12(4) & -1.24 & 0.06 & -0.01 & 0.10 & 0.23 & 0.05& main bulge \\
Djorg 2 & 8(2) & -1.15 & 0.05 & -0.01 & 0.08 & 0.33 & 0.03& main bulge \\
M4 & 205(68) & -1.13 & 0.10 & 0.29 & 0.10 & 0.33 & 0.05& Low-energy \\
M107 & 52(17) & -1.12 & 0.11 & 0.13 & 0.06 & 0.26 & 0.07& main bulge \\
NGC 6723 & 9(3) & -1.07 & 0.05 & 0.05 & 0.02 & 0.30 & 0.04&  main bulge\\
NGC 6316 & 8(2) & -0.89 & 0.01 & -0.08 & 0.11 & 0.28 & 0.02&  main bulge\\
NGC 6760 & 9(3) & -0.78 & 0.02 & 0.09 & 0.03 & 0.32 & 0.04&  main disk\\
47 Tuc & 270(90) & -0.78 & 0.05 & 0.22 & 0.04 & 0.34 & 0.03& main disk \\
M71 & 108(36) & -0.77 & 0.05 & 0.15 & 0.04 & 0.32 & 0.03& main disk \\
NGC 6642 & 11(3) & -0.64 & 0.69 & 0.04 & 0.06 & 0.22 & 0.14&  main bulge\\
NGC 6380 & 18(6) & -0.63 & 0.29 & 0.13 & 0.06 & 0.29 & 0.09&  main bulge\\
NGC 6388 & 34(11) & -0.54 & 0.28 & -0.09 & 0.09 & 0.08 & 0.04&  main bulge\\
NGC 6569 & 9(3) & -0.53 & 0.67 & 0.09 & 0.04 & 0.24 & 0.12&  main bulge\\
NGC 6304 & 13(4) & -0.51 & 0.09 & 0.14 & 0.07 & 0.28 & 0.06&  main bulge\\
NGC 6553 & 8(2) & -0.17 & 0.00 & 0.05 & 0.04 & 0.15 & 0.01&  main bulge\\
Terzan 5 & 9(3) & -0.09 & 0.47 & -0.02 & 0.03 & 0.12 & 0.13& main bulge \\
\enddata
\tablecomments{$^a$: total sample sizes of every GCs, and sample sizes of primordial populations in parentheses.  $^b$: mean abundances of primordial populations. $^c$: the associated standard deviations.
}
\end{deluxetable}

\bibliography{reference}{}

\begin{thebibliography}{}
\expandafter\ifx\csname natexlab\endcsname\relax\def\natexlab#1{#1}\fi
\providecommand{\url}[1]{\href{#1}{#1}}
\providecommand{\dodoi}[1]{doi:~\href{http://doi.org/#1}{\nolinkurl{#1}}}
\providecommand{\doeprint}[1]{\href{http://ascl.net/#1}{\nolinkurl{http://ascl.net/#1}}}
\providecommand{\doarXiv}[1]{\href{https://arxiv.org/abs/#1}{\nolinkurl{https://arxiv.org/abs/#1}}}

\bibitem[{M.~A. Beasley {et~al.}(2002)Beasley, Baugh, Forbes, Sharples, \& Frenk}]{beasley_formation_2002}
Beasley, M.~A., Baugh, C.~M., Forbes, D.~A., Sharples, R.~M., \& Frenk, C.~S. 2002, \bibinfo{title}{On the formation of globular cluster systems in a hierarchical {Universe},} Monthly Notices of the Royal Astronomical Society, 333, 383, \dodoi{10.1046/j.1365-8711.2002.05402.x}

\bibitem[{V. {Belokurov} \& A. {Kravtsov}(2024){Belokurov} \& {Kravtsov}}]{belokurov_-situ_2024}
{Belokurov}, V., \& {Kravtsov}, A. 2024, \bibinfo{title}{{In-situ versus accreted Milky Way globular clusters: a new classification method and implications for cluster formation},} \mnras, 528, 3198, \dodoi{10.1093/mnras/stad3920}

\bibitem[{V. {Belokurov} {et~al.}(2020){Belokurov}, {Sanders}, {Fattahi}, {Smith}, {Deason}, {Evans}, \& {Grand}}]{Belokurov2020}
{Belokurov}, V., {Sanders}, J.~L., {Fattahi}, A., {et~al.} 2020, \bibinfo{title}{{The biggest splash},} \mnras, 494, 3880, \dodoi{10.1093/mnras/staa876}

\bibitem[{I.~S. {Bowen} \& J. {Vaughan}(1973){Bowen} \& {Vaughan}}]{Bowen1973}
{Bowen}, I.~S., \& {Vaughan}, A.~H., J. 1973, \bibinfo{title}{{The optical design of the 40-in. telescope and of the Ir{\'e}n{\'e}e DuPont telescope at Las Campanas Observatory, Chile.},} \ao, 12, 1430, \dodoi{10.1364/AO.12.001430}

\bibitem[{J.~P. Brodie \& J. Strader(2006)Brodie \& Strader}]{brodie_extragalactic_2006}
Brodie, J.~P., \& Strader, J. 2006, \bibinfo{title}{Extragalactic {Globular} {Clusters} and {Galaxy} {Formation},} Annual Review of Astronomy and Astrophysics, 44, 193, \dodoi{10.1146/annurev.astro.44.051905.092441}

\bibitem[{T.~M. {Callingham} {et~al.}(2022){Callingham}, {Cautun}, {Deason}, {Frenk}, {Grand}, \& {Marinacci}}]{Callingham2022}
{Callingham}, T.~M., {Cautun}, M., {Deason}, A.~J., {et~al.} 2022, \bibinfo{title}{{The chemo-dynamical groups of Galactic globular clusters},} \mnras, 513, 4107, \dodoi{10.1093/mnras/stac1145}

\bibitem[{J.~L. {Carlin} {et~al.}(2018){Carlin}, {Sheffield}, {Cunha}, \& {Smith}}]{Carlin2018}
{Carlin}, J.~L., {Sheffield}, A.~A., {Cunha}, K., \& {Smith}, V.~V. 2018, \bibinfo{title}{{Chemical Abundances of Hydrostatic and Explosive Alpha-elements in Sagittarius Stream Stars},} \apjl, 859, L10, \dodoi{10.3847/2041-8213/aac3d8}

\bibitem[{E. {Carretta} {et~al.}(2009){Carretta}, {Bragaglia}, {Gratton}, {Lucatello}, {Catanzaro}, {Leone}, {Bellazzini}, {Claudi}, {D'Orazi}, {Momany}, {Ortolani}, {Pancino}, {Piotto}, {Recio-Blanco}, \& {Sabbi}}]{Carretta2009}
{Carretta}, E., {Bragaglia}, A., {Gratton}, R.~G., {et~al.} 2009, \bibinfo{title}{{Na-O anticorrelation and HB. VII. The chemical composition of first and second-generation stars in 15 globular clusters from GIRAFFE spectra},} \aap, 505, 117, \dodoi{10.1051/0004-6361/200912096}

\bibitem[{E. Ceccarelli {et~al.}(2025)Ceccarelli, Massari, Aguado-Agelet, Mucciarelli, Cassisi, Monelli, Pancino, Salaris, \& Saracino}]{ceccarelli_cluster_2025}
Ceccarelli, E., Massari, D., Aguado-Agelet, F., {et~al.} 2025, \bibinfo{title}{Cluster {Ages} to {Reconstruct} the {Milky} {Way} {Assembly} ({CARMA}). {III}. {NGC} 288 as the first {Splashed} globular cluster,} arXiv, \dodoi{10.48550/arXiv.2503.02939}

\bibitem[{Y. Chen \& O.~Y. Gnedin(2024)Chen \& Gnedin}]{chen_galaxy_2024}
Chen, Y., \& Gnedin, O.~Y. 2024, \bibinfo{title}{Galaxy assembly revealed by globular clusters,} The Open Journal of Astrophysics, 7, \dodoi{10.33232/001c.116169}

\bibitem[{P. Das {et~al.}(2020)Das, Hawkins, \& Jofré}]{das_ages_2020}
Das, P., Hawkins, K., \& Jofré, P. 2020, \bibinfo{title}{Ages and kinematics of chemically selected, accreted {Milky} {Way} halo stars,} MNRAS, 493, 5195, \dodoi{10.1093/mnras/stz3537}

\bibitem[{J.~G. {Fern{\'a}ndez-Trincado} {et~al.}(2021){Fern{\'a}ndez-Trincado}, {Minniti}, {Souza}, {Beers}, {Geisler}, {Moni Bidin}, {Villanova}, {Majewski}, {Barbuy}, {P{\'e}rez-Villegas}, {Henao}, {Romero-Colmenares}, {Roman-Lopes}, \& {Lane}}]{FT2021}
{Fern{\'a}ndez-Trincado}, J.~G., {Minniti}, D., {Souza}, S.~O., {et~al.} 2021, \bibinfo{title}{{VVV CL001: Likely the Most Metal-poor Surviving Globular Cluster in the Inner Galaxy},} \apjl, 908, L42, \dodoi{10.3847/2041-8213/abdf47}

\bibitem[{D.~A. Forbes \& T. Bridges(2010)Forbes \& Bridges}]{forbes_accreted_2010}
Forbes, D.~A., \& Bridges, T. 2010, \bibinfo{title}{Accreted versus \textit{in situ} {Milky} {Way} globular clusters,} MNRAS, 404, 1203, \dodoi{10.1111/j.1365-2966.2010.16373.x}

\bibitem[{E.~R. {Garro} {et~al.}(2022){Garro}, {Minniti}, {G{\'o}mez}, {Alonso-Garc{\'\i}a}, {Ripepi}, {Fern{\'a}ndez-Trincado}, \& {Vivanco C{\'a}diz}}]{Garro2022}
{Garro}, E.~R., {Minniti}, D., {G{\'o}mez}, M., {et~al.} 2022, \bibinfo{title}{{Inspection of 19 globular cluster candidates in the Galactic bulge with the VVV survey},} \aap, 658, A120, \dodoi{10.1051/0004-6361/202141819}

\bibitem[{D. Geisler {et~al.}(2021)Geisler, Villanova, O’Connell, Cohen, Moni~Bidin, Fernández-Trincado, Muñoz, Minniti, Zoccali, Rojas-Arriagada, Contreras~Ramos, Catelan, Mauro, Cortés, Ferreira~Lopes, Arentsen, Starkenburg, Martin, Tang, Parisi, Alonso-García, Gran, Cunha, Smith, Majewski, Jönsson, García-Hernández, Horta, Mészáros, Monaco, Monachesi, Muñoz, Brownstein, Beers, Lane, Barbuy, Sobeck, Henao, González-Díaz, Miranda, Reinarz, \& Santander}]{geisler_capos_2021}
Geisler, D., Villanova, S., O’Connell, J.~E., {et~al.} 2021, \bibinfo{title}{{CAPOS}: {The} bulge {Cluster} {APOgee} {Survey}: {I}. {Overview} and initial {ASPCAP} results,} Astronomy \& Astrophysics, 652, A157, \dodoi{10.1051/0004-6361/202140436}

\bibitem[{J.~E. {Gunn} {et~al.}(2006){Gunn}, {Siegmund}, {Mannery}, {Owen}, {Hull}, {Leger}, {Carey}, {Knapp}, {York}, {Boroski}, {Kent}, {Lupton}, {Rockosi}, {Evans}, {Waddell}, {Anderson}, {Annis}, {Barentine}, {Bartoszek}, {Bastian}, {Bracker}, {Brewington}, {Briegel}, {Brinkmann}, {Brown}, {Carr}, {Czarapata}, {Drennan}, {Dombeck}, {Federwitz}, {Gillespie}, {Gonzales}, {Hansen}, {Harvanek}, {Hayes}, {Jordan}, {Kinney}, {Klaene}, {Kleinman}, {Kron}, {Kresinski}, {Lee}, {Limmongkol}, {Lindenmeyer}, {Long}, {Loomis}, {McGehee}, {Mantsch}, {Neilsen}, {Neswold}, {Newman}, {Nitta}, {Peoples}, {Pier}, {Prieto}, {Prosapio}, {Rivetta}, {Schneider}, {Snedden}, \& {Wang}}]{Gunn2006}
{Gunn}, J.~E., {Siegmund}, W.~A., {Mannery}, E.~J., {et~al.} 2006, \bibinfo{title}{{The 2.5 m Telescope of the Sloan Digital Sky Survey},} \aj, 131, 2332, \dodoi{10.1086/500975}

\bibitem[{K. Hawkins {et~al.}(2015)Hawkins, Jofré, Masseron, \& Gilmore}]{hawkins_using_2015}
Hawkins, K., Jofré, P., Masseron, T., \& Gilmore, G. 2015, \bibinfo{title}{Using chemical tagging to redefine the interface of the {Galactic} disc and halo,} Mon. Not. R. Astron. Soc., 453, 758, \dodoi{10.1093/mnras/stv1586}

\bibitem[{D. {Horta} \& M.~K. {Ness}(2025){Horta} \& {Ness}}]{horta2025}
{Horta}, D., \& {Ness}, M.~K. 2025, \bibinfo{title}{{Hydrostatic and explosive $α$-element chemical abundances of Milky Way globular clusters, halo substructures, and satellite galaxies},} arXiv e-prints, arXiv:2506.08079, \dodoi{10.48550/arXiv.2506.08079}

\bibitem[{D. Horta {et~al.}(2020)Horta, Schiavon, Mackereth, Beers, Fernández-Trincado, Frinchaboy, García-Hernández, Geisler, Hasselquist, Jönsson, Lane, Majewski, Mészáros, Bidin, Nataf, Roman-Lopes, Nitschelm, Vargas-González, \& Zasowski}]{horta_chemical_2020}
Horta, D., Schiavon, R.~P., Mackereth, J.~T., {et~al.} 2020, \bibinfo{title}{The chemical compositions of accreted and \textit{in situ} galactic globular clusters according to {SDSS}/{APOGEE},} MNRAS, 493, 3363, \dodoi{10.1093/mnras/staa478}

\bibitem[{D. {Horta} {et~al.}(2021){Horta}, {Schiavon}, {Mackereth}, {Pfeffer}, {Mason}, {Kisku}, {Fragkoudi}, {Allende Prieto}, {Cunha}, {Hasselquist}, {Holtzman}, {Majewski}, {Nataf}, {O'Connell}, {Schultheis}, \& {Smith}}]{horta_evidence_2021}
{Horta}, D., {Schiavon}, R.~P., {Mackereth}, J.~T., {et~al.} 2021, \bibinfo{title}{{Evidence from APOGEE for the presence of a major building block of the halo buried in the inner Galaxy},} \mnras, 500, 1385, \dodoi{10.1093/mnras/staa2987}

\bibitem[{R. {Huang} {et~al.}(2024){Huang}, {Tang}, {Li}, {Geisler}, {Mateo}, {Song}, {Baumgardt}, {Carballo-Bello}, {Wang}, {Nie}, {Dias}, \& {Fern{\'a}ndez-Trincado}}]{HuangTang2024}
{Huang}, R., {Tang}, B., {Li}, C., {et~al.} 2024, \bibinfo{title}{{Driving factors behind multiple populations},} Science China Physics, Mechanics, and Astronomy, 67, 259513, \dodoi{10.1007/s11433-023-2332-5}

\bibitem[{R.~A. Ibata {et~al.}(1994)Ibata, Gilmore, \& Irwin}]{ibata_dwarf_1994}
Ibata, R.~A., Gilmore, G., \& Irwin, M.~J. 1994, \bibinfo{title}{A dwarf satellite galaxy in {Sagittarius},} Nature, 370, 194, \dodoi{10.1038/370194a0}

\bibitem[{T. {Je{\v{r}}{\'a}bkov{\'a}} {et~al.}(2018){Je{\v{r}}{\'a}bkov{\'a}}, {Zonoozi}, {Kroupa}, {Beccari}, {Yan}, {Vazdekis}, \& {Zhang}}]{Jerabkova2018}
{Je{\v{r}}{\'a}bkov{\'a}}, T., {Zonoozi}, A.~H., {Kroupa}, P., {et~al.} 2018, \bibinfo{title}{{Impact of metallicity and star formation rate on the time-dependent, galaxy-wide stellar initial mass function},} \aap, 620, A39, \dodoi{10.1051/0004-6361/201833055}

\bibitem[{P. {Jofr{\'e}} {et~al.}(2025){Jofr{\'e}}, {Aguilera-G{\'o}mez}, {Villarreal}, {Cubillos}, {Das}, {Hua}, {Yates}, {Silva}, {Vitali}, {Pe{\~n}a}, {Signor}, {Walsen}, {Tissera}, {Rojas-Arriagada}, {Johnston}, {Gilmore}, \& {Foley}}]{Jofre2025}
{Jofr{\'e}}, P., {Aguilera-G{\'o}mez}, C., {Villarreal}, P., {et~al.} 2025, \bibinfo{title}{{Studying stellar populations in Omega Centauri with phylogenetics},} arXiv e-prints, arXiv:2504.01813, \dodoi{10.48550/arXiv.2504.01813}

\bibitem[{H. Jönsson {et~al.}(2020)Jönsson, Holtzman, Prieto, Cunha, García-Hernández, Hasselquist, Masseron, Osorio, Shetrone, Smith, Stringfellow, Bizyaev, Edvardsson, Majewski, Mészáros, Souto, Zamora, Beaton, Bovy, Donor, Pinsonneault, Poovelil, \& Sobeck}]{jonsson_apogee_2020}
Jönsson, H., Holtzman, J.~A., Prieto, C.~A., {et~al.} 2020, \bibinfo{title}{{APOGEE} {Data} and {Spectral} {Analysis} from {SDSS} {Data} {Release} 16: {Seven} {Years} of {Observations} {Including} {First} {Results} from {APOGEE}-{South},} The Astronomical Journal, 160, 120, \dodoi{10.3847/1538-3881/aba592}

\bibitem[{C. {Kobayashi} {et~al.}(2006){Kobayashi}, {Umeda}, {Nomoto}, {Tominaga}, \& {Ohkubo}}]{K06}
{Kobayashi}, C., {Umeda}, H., {Nomoto}, K., {Tominaga}, N., \& {Ohkubo}, T. 2006, \bibinfo{title}{{Galactic Chemical Evolution: Carbon through Zinc},} \apj, 653, 1145, \dodoi{10.1086/508914}

\bibitem[{M.~G.~H. {Krause} {et~al.}(2016){Krause}, {Charbonnel}, {Bastian}, \& {Diehl}}]{Krause2016}
{Krause}, M. G.~H., {Charbonnel}, C., {Bastian}, N., \& {Diehl}, R. 2016, \bibinfo{title}{{Gas expulsion in massive star clusters?. Constraints from observations of young and gas-free objects},} \aap, 587, A53, \dodoi{10.1051/0004-6361/201526685}

\bibitem[{D.~R. {Law} \& S.~R. {Majewski}(2010){Law} \& {Majewski}}]{Law2010}
{Law}, D.~R., \& {Majewski}, S.~R. 2010, \bibinfo{title}{{Assessing the Milky Way Satellites Associated with the Sagittarius Dwarf Spheroidal Galaxy},} \apj, 718, 1128, \dodoi{10.1088/0004-637X/718/2/1128}

\bibitem[{S.~R. {Majewski} {et~al.}(2017){Majewski}, {Schiavon}, {Frinchaboy}, {Allende Prieto}, {Barkhouser}, {Bizyaev}, {Blank}, {Brunner}, {Burton}, {Carrera}, {Chojnowski}, {Cunha}, {Epstein}, {Fitzgerald}, {Garc{\'{\i}}a P{\'e}rez}, {Hearty}, {Henderson}, {Holtzman}, {Johnson}, {Lam}, {Lawler}, {Maseman}, {M{\'e}sz{\'a}ros}, {Nelson}, {Coung Nguyen}, {Nidever}, {Pinsonneault}, {Shetrone}, {Smee}, {Smith}, {Stolberg}, {Skrutskie}, {Walker}, {Wilson}, {Zasowski}, {Anders}, {Basu}, {Beland}, {Blanton}, {Bovy}, {Brownstein}, {Carlberg}, {Chaplin}, {Chiappini}, {Eisenstein}, {Elsworth}, {Feuillet}, {Fleming}, {Galbraith-Frew}, {Garc{\'{\i}}a}, {An{\'{\i}}bal Garc{\'{\i}}a-Hern{\'a}ndez}, {Gillespie}, {Girardi}, {Gunn}, {Hasselquist}, {Hayden}, {Hekker}, {Ivans}, {Kinemuchi}, {Klaene}, {Mahadevan}, {Mathur}, {Mosser}, {Muna}, {Munn}, {Nichol}, {O'Connell}, {Parejko}, {Robin}, {Rocha-Pinto}, {Schultheis}, {Serenelli}, {Shane}, {Silva Aguirre}, {Sobeck}, {Thompson}, {Troup}, {Weinberg}, \&
  {Zamora}}]{Majewski2017}
{Majewski}, S.~R., {Schiavon}, R.~P., {Frinchaboy}, P.~M., {et~al.} 2017, \bibinfo{title}{{The Apache Point Observatory Galactic Evolution Experiment (APOGEE)},} \aj, 154, 94, \dodoi{10.3847/1538-3881/aa784d}

\bibitem[{A.~C. {Mason} {et~al.}(2025){Mason}, {Schiavon}, {Kamann}, {Smith}, {Horta}, {Anguiano}, {Cunha}, {M{\'e}sz{\'a}ros}, {Majewski}, {O'Connell}, {Allende Prieto}, \& {Saracino}}]{Mason2025}
{Mason}, A.~C., {Schiavon}, R.~P., {Kamann}, S., {et~al.} 2025, \bibinfo{title}{{Chemical tagging with APOGEE, MUSE, and HST: constraints on the formation of $ω$ Centauri},} arXiv e-prints, arXiv:2504.06341, \dodoi{10.48550/arXiv.2504.06341}

\bibitem[{D. Massari {et~al.}(2019)Massari, Koppelman, \& Helmi}]{massari_origin_2019}
Massari, D., Koppelman, H.~H., \& Helmi, A. 2019, \bibinfo{title}{Origin of the system of globular clusters in the {Milky} {Way},} A\&A, 630, L4, \dodoi{10.1051/0004-6361/201936135}

\bibitem[{T. {Masseron} {et~al.}(2016){Masseron}, {Merle}, \& {Hawkins}}]{Masseron2016}
{Masseron}, T., {Merle}, T., \& {Hawkins}, K. 2016, \bibinfo{title}{{BACCHUS: Brussels Automatic Code for Characterizing High accUracy Spectra},} \doeprint{1605.004}

\bibitem[{A. {McWilliam} {et~al.}(2013){McWilliam}, {Wallerstein}, \& {Mottini}}]{McWilliam2013}
{McWilliam}, A., {Wallerstein}, G., \& {Mottini}, M. 2013, \bibinfo{title}{{Chemistry of the Sagittarius Dwarf Galaxy: A Top-light Initial Mass Function, Outflows, and the R-process},} \apj, 778, 149, \dodoi{10.1088/0004-637X/778/2/149}

\bibitem[{A.~P. {Milone} \& A.~F. {Marino}(2022){Milone} \& {Marino}}]{Milone2022}
{Milone}, A.~P., \& {Marino}, A.~F. 2022, \bibinfo{title}{{Multiple Populations in Star Clusters},} Universe, 8, 359, \dodoi{10.3390/universe8070359}

\bibitem[{D. {Minniti} {et~al.}(2017){Minniti}, {Geisler}, {Alonso-Garc{\'\i}a}, {Palma}, {Beam{\'\i}n}, {Borissova}, {Catelan}, {Clari{\'a}}, {Cohen}, {Contreras Ramos}, {Dias}, {Fern{\'a}ndez-Trincado}, {G{\'o}mez}, {Hempel}, {Ivanov}, {Kurtev}, {Lucas}, {Moni-Bidin}, {Pullen}, {Ram{\'\i}rez Alegr{\'\i}a}, {Saito}, \& {Valenti}}]{Minniti2017}
{Minniti}, D., {Geisler}, D., {Alonso-Garc{\'\i}a}, J., {et~al.} 2017, \bibinfo{title}{{New VVV Survey Globular Cluster Candidates in the Milky Way Bulge},} \apjl, 849, L24, \dodoi{10.3847/2041-8213/aa95b8}

\bibitem[{S. Mészáros {et~al.}(2020)Mészáros, Masseron, García-Hernández, Prieto, Beers, Bizyaev, Chojnowski, Cohen, Cunha, Dell'Agli, Ebelke, Fernández-Trincado, Frinchaboy, Geisler, Hasselquist, Hearty, Holtzman, Johnson, Lane, Lacerna, Longa-Peña, Majewski, Martell, Minniti, Nataf, Nidever, Pan, Schiavon, Shetrone, Smith, Sobeck, Stringfellow, Szigeti, Tang, Wilson, \& Zamora}]{meszaros_homogeneous_2020}
Mészáros, S., Masseron, T., García-Hernández, D.~A., {et~al.} 2020, \bibinfo{title}{Homogeneous {Analysis} of {Globular} {Clusters} from the {APOGEE} {Survey} with the {BACCHUS} {Code}. {II}. {The} {Southern} {Clusters} and {Overview},} MNRAS, 492, 1641, \dodoi{10.1093/mnras/stz3496}

\bibitem[{D.~L. {Nidever} {et~al.}(2015){Nidever}, {Holtzman}, {Allende Prieto}, {Beland}, {Bender}, {Bizyaev}, {Burton}, {Desphande}, {Fleming}, {Garc{\'{\i}}a P{\'e}rez}, {Hearty}, {Majewski}, {M{\'e}sz{\'a}ros}, {Muna}, {Nguyen}, {Schiavon}, {Shetrone}, {Skrutskie}, {Sobeck}, \& {Wilson}}]{Nidever2015}
{Nidever}, D.~L., {Holtzman}, J.~A., {Allende Prieto}, C., {et~al.} 2015, \bibinfo{title}{{The Data Reduction Pipeline for the Apache Point Observatory Galactic Evolution Experiment},} \aj, 150, 173, \dodoi{10.1088/0004-6256/150/6/173}

\bibitem[{S. Ortolani {et~al.}(1999)Ortolani, Bica, \& Barbuy}]{ortolani_blue_1999}
Ortolani, S., Bica, E., \& Barbuy, B. 1999, \bibinfo{title}{Blue horizontal branch globular clusters towardsthe bulge: {Terzan} 9, {NGC} 6139 and {NGC} 6453,} Astronomy and Astrophysics Supplement Series, 138, 267, \dodoi{10.1051/aas:1999275}

\bibitem[{G. {Pagnini} {et~al.}(2023){Pagnini}, {Di Matteo}, {Khoperskov}, {Mastrobuono-Battisti}, {Haywood}, {Renaud}, \& {Combes}}]{Pagnini2023}
{Pagnini}, G., {Di Matteo}, P., {Khoperskov}, S., {et~al.} 2023, \bibinfo{title}{{The distribution of globular clusters in kinematic spaces does not trace the accretion history of the host galaxy},} \aap, 673, A86, \dodoi{10.1051/0004-6361/202245128}

\bibitem[{E. Pancino {et~al.}(2017)Pancino, Romano, Tang, Tautvaišienė, Casey, Gruyters, Geisler, San~Roman, Randich, Alfaro, Bragaglia, Flaccomio, Korn, Recio-Blanco, Smiljanic, Carraro, Bayo, Costado, Damiani, Jofré, Lardo, De~Laverny, Monaco, Morbidelli, Sbordone, Sousa, \& Villanova}]{pancino_gaia_2017}
Pancino, E., Romano, D., Tang, B., {et~al.} 2017, \bibinfo{title}{The \textit{{Gaia}} -{ESO} {Survey}: {Mg}-{Al} anti-correlation in {iDR4} globular clusters⋆⋆⋆,} A\&A, 601, A112, \dodoi{10.1051/0004-6361/201730474}

\bibitem[{R.~P. Schiavon {et~al.}(2024)Schiavon, Phillips, Myers, Horta, Minniti, Allende Prieto, Anguiano, Beaton, Beers, Brownstein, Cohen, Fernández-Trincado, Frinchaboy, Jönsson, Kisku, Lane, Majewski, Mason, Mészáros, \& Stringfellow}]{schiavon_apogee_2024}
Schiavon, R.~P., Phillips, S.~G., Myers, N., {et~al.} 2024, \bibinfo{title}{The {APOGEE} value-added catalogue of {Galactic} globular cluster stars,} MNRAS, 528, 1393, \dodoi{10.1093/mnras/stad3020}

\bibitem[{B. {Tang} {et~al.}(2023){Tang}, {Zhang}, {Yan}, {Zhang}, {Carigi}, \& {Fern{\'a}ndez-Trincado}}]{Tang2023}
{Tang}, B., {Zhang}, J., {Yan}, Z., {et~al.} 2023, \bibinfo{title}{{Near-infrared chemical abundances of stars in the Sculptor dwarf galaxy},} \aap, 669, A125, \dodoi{10.1051/0004-6361/202244052}

\bibitem[{B. {Tang} {et~al.}(2017){Tang}, {Cohen}, {Geisler}, {Schiavon}, {Majewski}, {Villanova}, {Carrera}, {Zamora}, {Garcia-Hernandez}, {Shetrone}, {Frinchaboy}, {Meza}, {Fern{\'a}ndez-Trincado}, {Mu{\~n}oz}, {Lin}, {Lane}, {Nitschelm}, {Pan}, {Bizyaev}, {Oravetz}, \& {Simmons}}]{Tang2017}
{Tang}, B., {Cohen}, R.~E., {Geisler}, D., {et~al.} 2017, \bibinfo{title}{{Two groups of red giants with distinct chemical abundances in the bulge globular cluster NGC 6553 through the eyes of APOGEE},} \mnras, 465, 19, \dodoi{10.1093/mnras/stw2739}

\bibitem[{B. {Tang} {et~al.}(2018){Tang}, {Fern{\'a}ndez-Trincado}, {Geisler}, {Zamora}, {M{\'e}sz{\'a}ros}, {Masseron}, {Cohen}, {Garc{\'\i}a-Hern{\'a}ndez}, {Dell'Agli}, {Beers}, {Schiavon}, {Sohn}, {Hasselquist}, {Robin}, {Shetrone}, {Majewski}, {Villanova}, {Schiappacasse Ulloa}, {Lane}, {Minnti}, {Roman-Lopes}, {Almeida}, \& {Moreno}}]{Tang2018}
{Tang}, B., {Fern{\'a}ndez-Trincado}, J.~G., {Geisler}, D., {et~al.} 2018, \bibinfo{title}{{The Metal-poor non-Sagittarius (?) Globular Cluster NGC 5053: Orbit and Mg, Al, and Si Abundances},} \apj, 855, 38, \dodoi{10.3847/1538-4357/aaaaea}

\bibitem[{M. Unavane {et~al.}(1996)Unavane, Wyse, \& Gilmore}]{unavane_merging_1996}
Unavane, M., Wyse, R. F.~G., \& Gilmore, G. 1996, \bibinfo{title}{The merging history of the {Milky} {Way},} MNRAS, 278, 727, \dodoi{10.1093/mnras/278.3.727}

\bibitem[{M. {Xiang} \& H.-W. {Rix}(2022){Xiang} \& {Rix}}]{xiang2022}
{Xiang}, M., \& {Rix}, H.-W. 2022, \bibinfo{title}{{A time-resolved picture of our Milky Way's early formation history},} \nat, 603, 599, \dodoi{10.1038/s41586-022-04496-5}

\bibitem[{G. {Zasowski} {et~al.}(2017){Zasowski}, {Cohen}, {Chojnowski}, {Santana}, {Oelkers}, {Andrews}, {Beaton}, {Bender}, {Bird}, {Bovy}, {Carlberg}, {Covey}, {Cunha}, {Dell'Agli}, {Fleming}, {Frinchaboy}, {Garc{\'\i}a-Hern{\'a}ndez}, {Harding}, {Holtzman}, {Johnson}, {Kollmeier}, {Majewski}, {M{\'e}sz{\'a}ros}, {Munn}, {Mu{\~n}oz}, {Ness}, {Nidever}, {Poleski}, {Rom{\'a}n-Z{\'u}{\~n}iga}, {Shetrone}, {Simon}, {Smith}, {Sobeck}, {Stringfellow}, {Szigeti{\'a}ros}, {Tayar}, \& {Troup}}]{Zasowski2017}
{Zasowski}, G., {Cohen}, R.~E., {Chojnowski}, S.~D., {et~al.} 2017, \bibinfo{title}{{Target Selection for the SDSS-IV APOGEE-2 Survey},} \aj, 154, 198, \dodoi{10.3847/1538-3881/aa8df9}

\end{thebibliography}
\bibliographystyle{aasjournalv7}



\end{document}